\documentclass[%
superscriptaddress,
 amsmath,amssymb,
 aps, physrev,
twocolumn
]{revtex4-2}

\usepackage{graphicx}
\usepackage{dcolumn}
\usepackage{bm}
\usepackage{color}
\usepackage{siunitx}

\newcommand{\SLAC}{SLAC National Accelerator Laboratory, Menlo Park, CA 94025}
\newcommand{\AP}{Department of Applied Physics, Stanford University, Stanford, CA 94305}
\newcommand{\PULSE}{Stanford PULSE Institute, SLAC National Accelerator Laboratory, Menlo Park, CA 94025, USA}

\begin{document}

\preprint{APS/123-QED}

\title{\textbf{Broadband hard X-ray attosecond pulses from extremely chirped electron beams}} 

\author{River R. Robles}
\email{riverr@slac.stanford.edu}
\affiliation{\SLAC}
\affiliation{\AP}
\affiliation{\PULSE}

\author{Veronica Guo}
\affiliation{\SLAC}
\affiliation{\AP}
\affiliation{\PULSE}

\author{David Cesar}
\affiliation{\SLAC}

\author{Paris Franz}
\affiliation{\SLAC}
\affiliation{\AP}
\affiliation{\PULSE}

\author{Aliaksei Halavanau}
\affiliation{\SLAC}

\author{Alberto Lutman}
\affiliation{\SLAC}

\author{Takahiro Sato}
\affiliation{\SLAC}
\affiliation{\PULSE}

\author{Sanghoon Song}
\affiliation{\SLAC}

\author{Nicholas Sudar}
\affiliation{\SLAC}

\author{Yanwen Sun}
\affiliation{\SLAC}

\author{Zhen Zhang}
\affiliation{\SLAC}

\author{Diling Zhu}
\affiliation{\SLAC}

\author{Agostino Marinelli}
\email{marinelli@slac.stanford.edu}
\affiliation{\SLAC}
\affiliation{\PULSE}

\date{\today}

\begin{abstract}
Attosecond pulses from free-electron lasers have opened the doors to atomic site-specific studies of bound electronic dynamics on their natural, sub-femtosecond timescales. Key to their success has been electron beam shaping techniques enabling the generation of sub-femtosecond current spikes with peak currents on the order of 10 kA. We demonstrate in an RF linac the generation of current spikes with extreme chirps on the order of 350 MeV/$\mu$m, directly competitive with the chirps expected from beam-driven plasma wakefield accelerators. Leveraging chirp-taper compensation, we use these highly chirped beams to generate hard X-ray attosecond pulses with bandwidths exceeding 30 eV, a factor of two beyond previous demonstrations. We simultaneously present the first explicit experimental evidence of chirp-taper compensation in an attosecond XFEL, finding that optimal tapering improves the bandwidth and pulse energy by factors of two and five, respectively, for our conditions. In addition to the immediate utility of such broadband hard X-ray pulses, electron beams with such extreme chirps can be utilized for unique new experimental modalities by performing further compression after the undulators. Such post-lasing compression can enable subsequent superradiant light emission at longer wavelengths, or direct excitation of quantum systems with the beam's intense space-charge field for unique attosecond pump-probe possibilities. 
\end{abstract}

\maketitle

X-ray free-electron lasers (XFELs) leverage the resonant interaction between electrons and photons in an undulator to generate microbunching in the electron beam, enhancing the resulting radiation's peak power by many orders of magnitude compared to synchrotron radiation light sources \cite{bonifacio1984collective,pellegrini2016physics,huang2007review}. 
XFELs can produce attosecond pulses using electron beam shaping techniques to generate sub-femtosecond, 10 kA-scale current spikes \cite{marinelli2017experimental,huang2017generating,zhang2020experimental,duris2020tunable,malyzhenkov2020single,duris2021controllable,prat2023coherent,franz2024,yan2024terawatt,li2024beam,prat2025enhanced,guo2026high}. 
Soft X-ray attosecond science at FELs has matured rapidly in recent years thanks to synergistic developments in accelerator techniques \cite{duris2020tunable,duris2021controllable,zhang2020experimental,prat2023coherent,li2024beam,berrah2025attosecond} and X-ray diagnostics \cite{hartmann2018attosecond,li2018characterizing}.
Hard X-ray attosecond science, on the other hand, has until recently been limited by a lack of pulse duration measurements and limited broadband X-ray optics for characterizing and manipulating hard X-ray attosecond pulses. 
Previous experiments have employed nonlinear compression \cite{huang2017generating,malyzhenkov2020single,yan2024terawatt,prat2025enhanced} and emittance spoiling \cite{marinelli2017experimental} to generate single spike hard X-ray pulses with bandwidths as high as 15 eV, sufficiently broad to support attosecond pulses, though with no actual pulse duration measurement. 

A recent experiment by Inoue \textit{et al} leveraged well-characterized nonlinear processes to provide the first direct evidence of hard X-ray attosecond pulses \cite{inoue2025experimental}. 
This experiment benefited from significant recent advances in broadband hard X-ray optics. 
Advanced Kirkpatrick-Baez mirror systems have enabled focusing of XFEL pulses to nanometer spots \cite{yamada2024extreme}, including attosecond hard X-ray pulses \cite{inoue2025nanofocused}. 
Furthermore, a hard X-ray attosecond split and delay device was recently designed \cite{sun2025ultrastable} and commissioned \cite{sun2026hard} with attosecond stability and tunability, enabling hard X-ray attosecond pump-probe methodologies.
These recent achievements promise to promote hard X-ray attosecond pulses from a topic of accelerator research into robust application, including ultrafast scattering experiments and electronic-damage-free measurements in the hard X-ray regime. 
This rapid recent progress demands a reexamination of the physics and techniques enabling the generation of hard X-ray attosecond pulses, including their ultimate limiting factors.

Attosecond XFELs are enabled by a sub-femtosecond, high current spike in the electron beam that both shortens the FEL cooperation length and restricts lasing to a short region around the spike \cite{bonifacio1994spectrum,zholents2004proposal,zholents2005method}. 
These high current spikes experience very strong collective forces, especially longitudinal space-charge \cite{ding2009generation,geloni2007longitudinal}.
Those collective effects inevitably lead to large energy modulations, inducing complex sub-femtosecond phase space structures, especially a very strong linear chirp in the center of the current spike. 
A linear chirp generally degrades FEL amplification \cite{HuangWKB}, but can be compensated for by tapering the undulators in a scheme called chirp-taper compensation \cite{saldin2006self,giannessi2011self}. 
The concept of chirp-taper compensation was first demonstrated at visible wavelengths \cite{giannessi2011self,marcus2012time}. Although undulator tapering is routinely applied to attosecond XFEL generation, chirp-taper compensation has never been directly demonstrated experimentally in the X-ray regime, and its impact on the pulse properties has never been quantified in the scientific literature.
Such experimental demonstration is important, as chirp-taper compensation lies at the heart of a wide variety of proposed FEL techniques aiming to generate short pulses with high peak powers \cite{tanaka2015proposal,prat2015simple,prat2015efficient,tanaka2016using,duris2020superradiant,tang2022electron,schneidmiller2022application,liu2025generation,xu2026high}, multiple pulses \cite{zhang2019double,sun2024attosecond}, pulses with programmable spectrotemporal structure \cite{xu2025attosecond,rebernik2026few}, or leveraging the naturally chirped beams from plasma wakefield accelerators for FEL lasing \cite{maier2012demonstration,teng2026phase}.
At the same time, generating large energy chirps is essential for proposed extreme compression schemes with applications to superradiant light emission \cite{tibai2014proposal,toth2018single,tibai2018carrier,mak2019attosecond,rebernik2020isolated,emma2021terawatt,hessami2024wavelength,swanson2025longitudinal} and direct driving of quantum dynamics \cite{cesar2022ultrafast,kim2026coherent}.

\begin{figure*}[htb]
    \centering
    \includegraphics{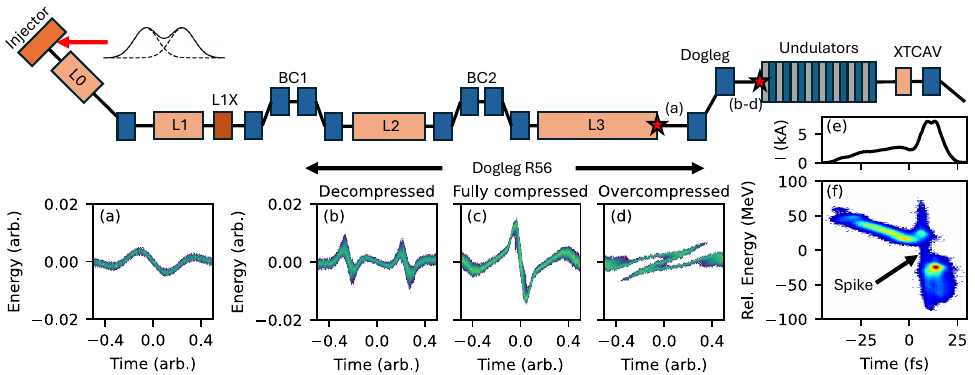}
    \caption{Schematic of the LCLS accelerator complex, showing a cartoon of the shaped drive laser pulse. (a) A simplified illustration of the kind of single cycle energy modulation that can form after the third linac. (b)-(d) The longitudinal phase space after compression in the dogleg and subsequent modulation by collective effects for three different $R_{56}$ working points. (e)-(f) The measured current profile and longitudinal phase space after the undulators. In these and other longitudinal phase space plots in the paper, the head of the beam lies to the left.}
    \label{fig:schematic}
\end{figure*}

In this article, we use photocathode laser shaping \cite{zhang2020experimental} to generate current spikes with extremely large energy chirps on the order of 350 MeV/$\mu$m. 
We show that compensating this chirp with undulator tapering leads to the broadest single spike hard X-ray bandwidths reported to date: more than 30 eV. 
We verify the production of such intense chirps by leveraging chirp-taper compensation as a diagnostic. 
We show first that a controllable current spike can be generated which lases efficiently in spite of very large, 100 MeV energy spreads. 
By scanning the taper, aided by X-ray pulse energy and spectrum measurements, we show direct evidence of chirp-taper compensation.
This constitutes the first explicit demonstration of chirp-taper compensation in the X-ray regime, and in attosecond XFELs in particular.
Optimal compensation increases the pulse energy by a factor of $\sim5$, and the bandwidth by a factor of two relative to the untapered case. 
The resulting $\sim$30 eV bandwidth exceeds that of earlier single spike hard X-ray demonstrations by at least a factor of two \cite{huang2017generating,marinelli2017experimental,malyzhenkov2020single,yan2024terawatt,prat2025enhanced}. 
That the broadband pulses generated by this method are sub-femtosecond was recently confirmed in the first experiments measuring hard X-ray attosecond pulse durations \cite{inoue2025experimental,sun2026hard}. 
Furthermore, the estimated chirp within the current spike is on the order of 350 MeV/$\mu$m, comparable to the chirps predicted at the output of beam-driven plasma wakefield accelerators (PWFAs) \cite{emma2021terawatt,swanson2025longitudinal}. 
With this work, we show that such extreme beams are in fact available today at existing facilities, and with sufficiently high brightness to support hard X-ray lasing. 
Such intense chirps can be compressed downstream of the undulators \cite{cesar:ipac24-mopg53}, yielding ultrashort current spikes that could be used to emit intense, carrier-envelope-phase (CEP) stable, superradiant pulses with intrinsic time synchronicity with the X-rays \cite{tibai2014proposal,hessami2024wavelength,toth2018single,tibai2018carrier,mak2019attosecond,rebernik2020isolated}, or whose space-charge fields can be used for direct excitation of quantum systems \cite{cesar2022ultrafast,kim2026coherent}.

An electron beam with current $I(s)$, Lorentz factor $\gamma$, transverse size $r_b$, and length $\sigma_z$ generates a longitudinal self-field of the form \cite{geloni2007longitudinal, ding2009generation}:
\begin{align}\label{eqn:LSC_field}
    E_z(s)\simeq -\frac{Z_0I'(s)}{4\pi\gamma^2}\left[2\ln\left(\frac{\gamma\sigma_z}{r_b}\right)+1\right],
\end{align}
where $Z_0$ is the free space impedance, and $s=z-v_bt$ is the longitudinal position. Stronger forces are generated in an undulator, where the same expression applies by replacing $\gamma$ with $\gamma_z=\gamma/\sqrt{1+K^2/2}$. $K$ is the peak undulator strength parameter, typically between 1 and 10 \cite{geloni2007longitudinal}. At the center of a Gaussian current spike, this leads to a roughly linear chirp with sufficient total energy spread to suppress FEL lasing. However, because the energy spread is correlated, it can be compensated by undulator tapering. For a beam with chirp $\frac{d\gamma}{ds}$, the appropriate taper rate is \cite{saldin2006self}
\begin{align}\label{eqn:chirp_taper_comp}
    \frac{dK}{dz} = \frac{2+K^2}{\gamma K}\frac{\lambda_r}{\lambda_u}\frac{d\gamma}{ds}.
\end{align}
Space-charge forces generate positive chirps, and therefore require positive undulator tapering. 

We performed our experiments at the Linac Coherent Light Source (LCLS). Figure~\ref{fig:schematic} shows the layout of the accelerator complex. The LCLS is an XFEL driven by a normal conducting copper linear accelerator which during this experiment delivered beams with 10.263 GeV average energy, sub-micron emittance, and MeV-level uncorrelated energy spread. The machine is driven by a normal conducting RF photoinjector, where the electron beam is generated through photoemission from a copper photocathode by a drive laser. 
At the end of the accelerator, the beam is transported through a magnetic dogleg into the undulators. The X-ray pulse energy and spectrum are measured using a gas detector and a bent crystal spectrometer, respectively. Additionally, an X-band transverse deflecting cavity (XTCAV) combined with a bending magnet enables imaging of the longitudinal phase space of the beam \cite{emma2000transverse,akre2002bunch,behrens2014few} with few femtosecond temporal resolution. 

Electron beam shaping to generate attosecond pulses is accomplished by shaping the photocathode drive laser \cite{zhang2020experimental}. At the LCLS, the nominal drive laser pulse is composed of two Gaussians stacked with a temporal offset such that the overall profile is a few ps long flat top. Small perturbations on these initial profiles can be amplified in a process known as the microbunching instability \cite{borland2002start,saldin2002klystron,saldin2004longitudinal,heifets2002coherent}. We seed the microbunching instability to generate a strong current spike in the beam by increasing the time delay between the two pulses to introduce a dip in the beam current profile (see Fig.~\ref{fig:schematic}, top left). The small initial current modulation leads to modulations of the longitudinal phase space through cascades of collective effects and bunch compression through the accelerator.

At the end of the accelerator, one final round of compression can be achieved by tuning quadrupole magnets in the dispersive dogleg \cite{nosochkov2021tuning}. This allows the generation of a tunable longitudinal dispersion $R_{56}$, resulting in a  relative longitudinal displacement of the particles by an amount:
\begin{equation}
    \Delta s=R_{56}\eta,
\end{equation}
where $\eta = \Delta\gamma/\gamma$ is the normalized energy deviation of a particle with respect to the nominal beam energy.
Given the strong positive energy chirp induced by self-forces, a negative $R_{56}$ compresses the electron bunch, while a positive value de-compresses it. After the dogleg, the beam is transported for 200 meters to the undulator entrance, during which collective effects can induce further energy modulation and chirp. 

To better illustrate the physics at play, Figure~\ref{fig:schematic}(a) shows an illustration of the kind of single cycle modulation that may form by the end of the last linac. Panels (b)-(d) show what happens to such a modulation after compression with a variable $R_{56}$ and subsequent modulation by collective effects. In panels (b) and (d), the modulation is either decompressed with a positive $R_{56}$ or overcompressed with a strong negative $R_{56}$. In panel (c), the initial modulation is fully compressed in the dogleg, yielding an isolated high current spike. As the beam propagates to the undulators, a strong linear chirp develops with significant total energy spread. Panels (e) and (f) show an experimentally measured current profile and longitudinal phase space, respectively, at the end of the accelerator. While the overall bunch duration is more than 50 fs, near the back of the beam there is a current spike which, through collective effects, has generated a large energy modulation. The peak to peak energy spread exceeds 100 MeV, greater than 1\% of the average energy and significantly larger than the energy spread typically allowed by the FEL process.

\begin{figure}[h!]
    \centering
    \includegraphics[width=\linewidth]{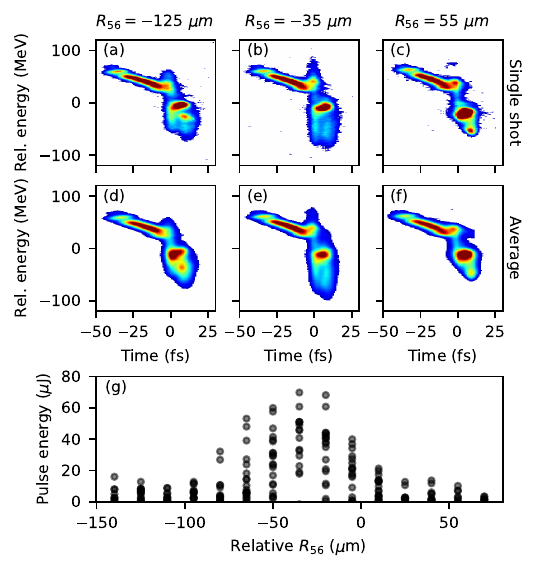}
    \caption{Summary of a scan of the dogleg $R_{56}$. (a)-(c) Single shot phase space images for three different values of $R_{56}$. (d)-(f) Phase space images averaged over twenty shots for the same values of $R_{56}$. (g) Pulse energy emitted at 9 keV as a function of the $R_{56}$ with the undulators linearly tapered to compensate a 400 MeV/$\mu$m chirp.}
    \label{fig:r56_scan}
\end{figure}

\begin{figure*}[htb]
    \centering
    \includegraphics[width=\linewidth]{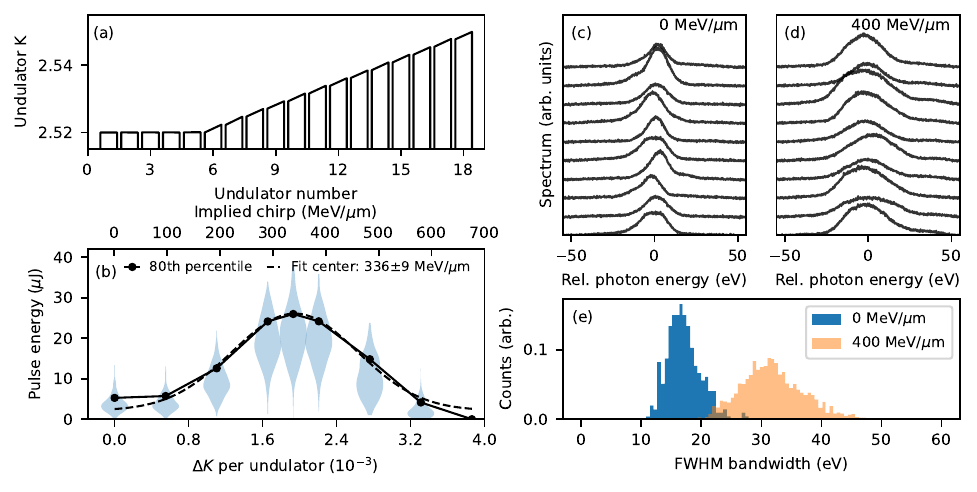}
    \caption{Summary of a taper scan for fixed dogleg $R_{56}$. (a) The undulator taper profile -- as the taper is scanned, the slope of the tapered section changes. (b) Pulse energy emitted at 9 keV as a function of the linear taper, peaking at the corresponding chirp of 333 MeV/$\mu$m. (c) and (d) show single shot spectral measurements at 9 keV for a flat taper and for a taper compensating a 400 MeV/$\mu$m chirp, respectively. (e) Histograms of the bandwidth generated in the two cases in (c) and (d). }
    \label{fig:nc_taper_scan}
\end{figure*}

In Figure~\ref{fig:r56_scan} we show an experimental scan of the dogleg $R_{56}$. Panels~(a)-(f) show the final longitudinal phase space measured after the undulators for three different $R_{56}$ values. Panels (a)-(c) show single shot data, while (d)-(f) show averages over 20 independent shots. We recall that any modulation on the beam at this point should be compressible by a negative $R_{56}$ if it derives from space-charge forces. From right to left, we can see a large growth of the energy spread in a narrow slice in the center of the beam as we pass through $R_{56}=-35$ $\mu$m. For the  positive $R_{56}$ we expect that the modulation is still undercompressed, while for the more negative one it should be overcompressed. The energy spread growth at $R_{56}=-35$ $\mu$m is evidence of generation of a large current spike in the middle of the beam, which generates large collective forces. Panel (g) shows the X-ray pulse energy emitted at 9 keV central photon energy as a function of the relative $R_{56}$. During the scan, the undulators were positively tapered at a rate corresponding to a chirp of 400 MeV/$\mu$m. The emitted pulse energy peaks at the same value of $R_{56}$ that maximizes the growth in energy spread in the XTCAV. This is consistent with the energy spread coming from the formation of a high current spike, and the resulting correlated energy spread being compensated by the undulator taper. 

We directly demonstrate chirp-taper compensation by scanning the taper rate while synchronously measuring the emitted pulse energy and the X-ray spectrum for $R_{56}=-35$ $\mu$m. Figure~\ref{fig:nc_taper_scan}(a) shows the undulator taper configuration used in the experiment. The current LCLS undulators are divided into 3.4 meter long segments consisting of 130 periods with a period of $2.6$ cm. The undulator segments are separated by 60 cm which contain matching quadrupoles, orbit correcting dipole magnets, and phase shifters. After five untapered undulators, twelve undulators are given a positive linear taper which we will quantify in terms of the change in $K$ per undulator section. Figure~\ref{fig:nc_taper_scan}(b) shows the distribution of emitted pulse energies, evaluated over 100 independent shots, as a function of the linear taper imposed in the undulators. We include two x axes. One gives the taper rate directly in units of $\Delta K$ per undulator, while the other gives it in terms of an implied chirp -- the chirp theoretically compensated by this taper according to Eqn.~\eqref{eqn:chirp_taper_comp}. With zero taper, only a few $\mu$J are emitted on average. As we scan the taper, we observe an increasing trend until we reach a taper of $2\times 10^{-3}$,  corresponding to a 333 MeV/$\mu$m chirp. For this taper, the 80th percentile pulse energy has increased to over 25 $\mu$J. Further increasing the taper suppresses lasing to negligible levels. Panels (c) and (d) show characteristic X-ray spectra measured later in the experiment after additional tuning for a flat taper and for a 400 MeV/$\mu$m matched taper, respectively. With no taper, single spike pulses are emitted with FWHM bandwidths around 18 eV (see panel (c)), sufficiently broad to support attosecond pulses though with very low pulse energy. At the 400 MeV/$\mu$m matched taper, the pulses maintain their single spike character, and the bandwidth is also significantly broader, roughly 30 eV on average, as shown in panel (d). For comparison, panel (e) shows histograms of the FWHM bandwidth for the untapered and tapered working points. Previous demonstrations of single spike hard X-ray pulses achieved bandwidths around 15 eV, so this represents a factor of two improvement in the available coherent bandwidth of single spike hard X-ray FEL pulses. The attosecond duration of these pulses has been confirmed in recent experiments \cite{inoue2025experimental,sun2026hard}. Furthermore, FEL simulations presented in Appendix~\ref{app:sims}, suggest that the broadened bandwidth corresponds directly with shorter pulse durations, not just an increased X-ray chirp.

The estimated $333$ MeV/$\mu$m chirp is very large, competitive with schemes using beam-driven PWFAs to intentionally generate hundreds of MeV/$\mu$m chirps \cite{emma2021terawatt,swanson2025longitudinal}. For example, the study in \cite{emma2021terawatt} shows PWFA simulations generating a 3.3 GeV beam with 3.7\%/$\mu$m chirp. Here we have experimentally demonstrated, using beam shaping techniques in a conventional RF linac, a 3.2\%/$\mu$m chirp, competitive with the simulated PWFA case. Furthermore, the observation of FEL lasing is evidence that the current spike maintains a low emittance and low slice energy spread. The chirp can be compressed with a small compressor \cite{cesar:ipac24-mopg53}, with $|R_{56}|\simeq\frac{1}{3.2\text{ \%/$\mu$m}}\simeq 30$~$\mu$m. The compressed spike length is limited by $R_{56}\sigma_{\eta,s}$, with $\sigma_{\eta,s}$ the relative slice energy spread. A beam capable of hard X-ray lasing must have $\sigma_{\eta,s}$ on the order of $10^{-4}-10^{-3}$, so $R_{56}\sigma_{\eta,s}$ can be as short as a few nanometers. Lasing to saturation increases the slice energy spread, but still only to the order of $10^{-3}$. Such a short current spike could emit high power, CEP stable attosecond pulses in the extreme ultraviolet range through superradiant emission in a short undulator \cite{tibai2014proposal,emma2021terawatt,hessami2024wavelength}. The resulting pulse would be naturally time synchronized with the X-rays emitted upstream, making for a uniquely powerful pump-probe experimental setup. Alternatively, the strong space-charge field can be used to directly excite quantum systems \cite{cesar:ipac24-mopg53,cesar2022ultrafast}. This result reveals that experiments which were previously thought to require reliable, beam-driven PWFA systems may in fact be possible today at existing facilities, with the benefit that the diagnostics to use these electron beams and X-ray pulses already exists in part in the XFEL user end stations. 

In conclusion, we have demonstrated the generation of electron beam current spikes with extremely large chirps on the order of 350 MeV/$\mu$m. In spite of the magnitude of the chirp, the beam maintains sufficiently high quality to generate single spike hard X-ray attosecond pulses with unprecedented bandwidths in excess of 30 eV. We show the first explicit evidence for chirp-taper compensation in the X-ray regime, finding that while single spike lasing was still possible without tapering, tapering both enhanced the pulse energy and significantly broadened the bandwidth. The emitted pulses have immediate applications in ultrafast hard X-ray science. Furthermore, the highly chirped current spike holds promise for future advanced attosecond pump-probe experiments. In that setting, further compression of the current spike can allow it to superradiantly emit a second short radiation pulse, or to directly drive a quantum system with the intense space-charge field of the beam, opening the doors to unique new experimental possibilities. 

\textit{Acknowledgements} -- This work was supported by the U.S. Department of Energy, Office of Science, Office of Basic Energy Sciences under
Contract No. DE-AC02-76SF00515, and by the U.S. Department of Energy, Office of Science, Office of Basic Energy Sciences Accelerator and Detector Research Program. R. R. R. acknowledges the support of the Robert H. Siemann Fellowship. 

\appendix
\section{Supporting simulations}\label{app:sims}

The bandwidth broadening observed when the undulators are optimally tapered could be attributed either to a direct shortening of the X-ray pulse duration or an increase in the X-ray chirp. To understand which option is the case here, we have performed FEL simulations using GENESIS 1.3 v4 \cite{reiche1999genesis,genesis4github}. The simulation parameters are specified in Table~\ref{tab:sim_parameters}. We simulate a short Gaussian beam, emulating only the current spike and ignoring the background current distribution for the sake of capturing the basic physics of chirp-taper compensation in the spike. We generate the beam with a purely linear chirp, which is an oversimplification of the true scenario but sufficient for demonstration.

\begin{table}[h!]
    \centering
    \begin{tabular}{c|c}
        \hline
        Beam parameter & Value  \\
        \hline
        Peak current & 20 kA\\
        RMS spike length & 60 nm (200 as)\\
        Chirp & 350 MeV/$\mu$m \\
        Beta function & 30 m\\
        Norm. emittance & 0.8 $\mu$m rad\\
        Energy & 10 GeV \\
        Uncorr. energy spread & 2 MeV\\
        \hline
        Undulator parameter & Value\\
        \hline
        Period & 2.6 cm \\
        Length & 32.5 m\\
        Resonant photon energy & 10 keV
    \end{tabular}
    \caption{Parameters for simulations.}
    \label{tab:sim_parameters}
\end{table}

\begin{figure}[h!]
    \centering
    \includegraphics[width=\linewidth]{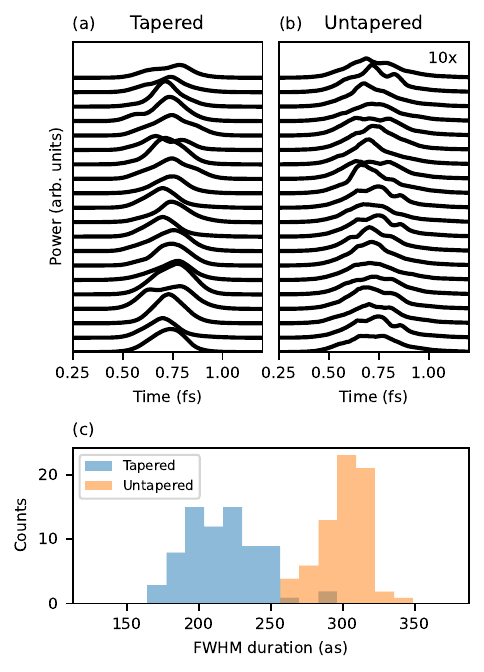}
    \caption{Summary of FEL simulation results. Panels (a) and (b) show 20 random power profiles from the tapered and untapered simulations, respectively. Panel (c) shows the distribution of pulse durations for the two cases. The tapering is seen to indeed shorten the pulse duration compared to the untapered case. }
    \label{fig:sim_results}
\end{figure}

We then performed two sets of simulations: one in which the undulators are not tapered, and one in which they are tapered at the ideal rate calculated from Eq.~\eqref{eqn:chirp_taper_comp}. For each case, we perform 100 simulations each starting from different random seeds for the initial shot noise microbunching. Figure~\ref{fig:sim_results} summarizes the simulation results. Panels (a) and (b) show 20 randomly sampled power profiles from the two simulations with and without tapering, respectively. Note that the profiles in the untapered case are multiplied by a factor of 10 to appear on the same scale as the tapered case, reflecting the significant power enhancement with proper tapering. Panel (c) shows histograms of the pulse durations for the two cases for pulses whose power is well-fit by a Gaussian ($r^2>0.98$). 74 out of 100 pulses satisfy this filter for the tapered case compared to 70 for the untapered case. The average pulse duration without tapering is around 300 as, while for the tapered case it is close to 200 as. This provides evidence that the bandwidth broadening we observe in the experiment reflects a shorter pulse duration, not just additional X-ray chirp, but this should eventually be directly confirmed by measurements. 

\bibliography{apssamp}

\end{document}